\documentclass[aps,prl,reprint,superscriptaddress]{revtex4-1}

\usepackage{xcolor}
\usepackage{graphicx}

\begin{document}


\title{Generalized models for bond percolation transitions of associative polymers}

\author{Jeong-Mo Choi}
\affiliation{Department of Biomedical Engineering, Washington University in St. Louis, St. Louis, MO 63130, United States}
\affiliation{Center for Science and Engineering of Living Systems (CSELS), Washington University in St. Louis, St. Louis, MO 63130, United States}
\affiliation{Natural Science Research Institute, Korea Advanced Institute of Science and Technology (KAIST), Yuseong-gu, Daejeon 31414, Republic of Korea}
\author{Anthony A. Hyman}
\affiliation{Max-Planck-Institut f\"ur Zellbiologie und Genetik, Pfotenhauerstra{\ss}e 108, 01307 Dresden, Germany}
\affiliation{Center for Systems Biology Dresden, Pfotenhauerstra{\ss}e 108, 01307 Dresden, Germany}
\author{Rohit V. Pappu}
\email[]{pappu@wustl.edu}
\affiliation{Department of Biomedical Engineering, Washington University in St. Louis, St. Louis, MO 63130, United States}
\affiliation{Center for Science and Engineering of Living Systems (CSELS), Washington University in St. Louis, St. Louis, MO 63130, United States}

\date{\today}

\begin{abstract}
Associative polymers with sticker-and-spacer architectures undergo phase transitions that combine phase separation and bond percolation. Here, we generalize mean-field models for bond percolation and extract effective sticker interaction strengths from experimental data for specific proteins. We use graph-based Monte Carlo simulations to evaluate effects ignored by mean-field models. These include bond cooperativity defined as changes to strengths of new bonds (physical crosslinks) given the presence of a prior bond. We find that bond cooperativity fundamentally alters phase behavior, especially in sub-stoichiometric regimes for valencies of complementary stickers.
\end{abstract}


\maketitle

Multivalent proteins and nucleic acids with multiple attractive groups (stickers) are biological instantiations of associative polymers \cite{rubinstein1997} that can be modeled using a \emph{stickers-and-spacers framework} \cite{doi:10.1021/ma970616h,doi:10.1021/acs.macromol.8b00661}. This framework has been used to explain, model, and design \emph{phase-separation} aided \emph{bond percolation} (PSBP) transitions of linear multivalent proteins \cite{WANG2018688,Martin694,doi:10.1146/annurev-biophys-121219-081629} that drive the formation of network-fluid-like condensates in living cells \cite{Shineaaf4382,doi:10.1146/annurev-genet-112618-043527}. Bond percolation in associative polymers is enabled by reversible physical crosslinks among stickers. This gives rise to system-specific percolation threshold concentrations ($c_{\text{perc}}$) that are governed by the valence (number) of stickers and sticker-sticker interaction strengths \cite{rubinstein1997,doi:10.1021/ma970616h}. The excluded volumes of spacers and weaker sticker-spacer as well as spacer-spacer attractions determine whether or not percolation is enabled by phase separation \cite{10.7554/eLife.30294}. In a binary mixture comprising associative polymers and solvent molecules, phase separation occurs above a saturation concentration ($c_{\text{sat}}$) where the system separates into dense, polymer-rich condensates that coexist with a solvent-rich dilute phase giving rise to a condensate-spanning percolated network \cite{10.1371/journal.pcbi.1007028}. In multivalent proteins, phase separation and percolation are likely to be strongly coupled and one can use $c_{\text{perc}}$ as a proxy for $c_{\text{sat}}$ \cite{rubinstein1997,doi:10.1021/ma970616h,10.1371/journal.pcbi.1007028}

Here, we assess how the interplay between homotypic and heterotypic sticker-sticker interactions influences $c_{\text{perc}}$ for associative polymers with more than two types of stickers. To do so, we generalize previous mean-field models to systems with arbitrary numbers of sticker types. Additionally, we use graph-based Monte Carlo simulations to show that the entropic cost associated with the incorporation of polymers into a growing network and the inclusion of bond cooperativity cause quantitative and testable departures from predictions based on mean-field calculations.

\section{A generalized mean-field model and its application}
A volume $V$ consists of $N$ associative polymers, each of which contains $n_i$ stickers of type $i$. A physical crosslink between sticker pairs of type $i$ and type $j$ leads to an energy gain of $\epsilon_{ij}$ and this interaction constrains the sticker pairs to a volume $v_{ij}$, which is the bond volume of the $i$-$j$ pair. Generalization of our previous approach \cite{WANG2018688} provides an expression for the percolation threshold \footnote{Details of the main derivation are in the \textbf{Supplemental Material}}:
\begin{equation} \label{eq:MFT}
c_{\text{perc}} = \frac{N}{V} \approx \frac{1}{\sum_i \lambda_{ii} n_i^2 + 2 \sum_{i, j > i} \lambda_{ij} n_i n_j},
\end{equation}
where $\lambda_{ij} = v_{ij} e^{-\beta \epsilon_{ij}}$ and $\beta = 1/(k_{\text{B}}T)$ is the inverse temperature. \textbf{Equation \ref{eq:MFT}} helps quantify the relative contributions of different types of stickers, including the interplay between homotypic and heterotypic interactions to $c_{\text{perc}}$.

Wang \emph{et al.} investigated PSBP transitions of several multivalent proteins including the protein FUS  that has an intrinsically disordered N-terminal prion-like domain (PLD) and C-terminal RNA binding domains (RBD) \cite{WANG2018688}. Saturation concentrations were found to be governed by the numbers of tyrosine (Tyr) residues in PLDs and arginine (Arg) residues in RBDs. Experiments show that the $c_{\text{sat}}$ for full-length FUS (FL-FUS) is 2 $\mu$M at 75 mM KCl, while its PLD as an independent module has an estimated $c_{\text{sat}}$ that is at least 120 $\mu$M \cite{BURKE2015231,WANG2018688}. The numbers of Tyr and Arg residues in FL-FUS are 33 and 36, respectively, and while the FUS PLD contains 27 Tyr residues, it has zero Arg residues. Based on our generalized mean-field model, the ratio of saturation concentrations for FL FUS and FUS PLD can be approximated as
\begin{equation} \label{eq:ratio}
\frac{c_{\text{sat}}\text{(FL FUS)}}{c_{\text{sat}}\text{(FUS PLD)}} \approx \frac{(n_{\text{Y}}^{\text{PLD}})^2}{(n_{\text{Y}}^{\text{FL}})^2 + 2 (\lambda_{\text{YR}}/\lambda_{\text{YY}}) n_{\text{Y}}^{\text{FL}} n_{\text{R}}^{\text{FL}}},
\end{equation}
where $n_{\text{Y}}^{\text{PLD}}$, $n_{\text{Y}}^{\text{FL}}$, and $n_{\text{R}}^{\text{FL}}$ refer to the numbers of Tyr residues in FUS PLD, Tyr residues in FL FUS, and Arg residues in FL FUS, respectively. Assuming the bond volumes to be identical, our analysis of the relative saturation concentrations shows that an interaction energy difference of $\epsilon_{\text{YY}} - \epsilon_{\text{YR}} = 0.11 k_{\text{B}} T$ between the Tyr-Tyr pair and the Tyr-Arg pair combined with multivalence is sufficient to account for the 100-fold increase of the saturation concentration of the PLD vis-\'{a}-vis FL FUS.

\section{Going beyond mean-field models}
\begin{figure}
\includegraphics{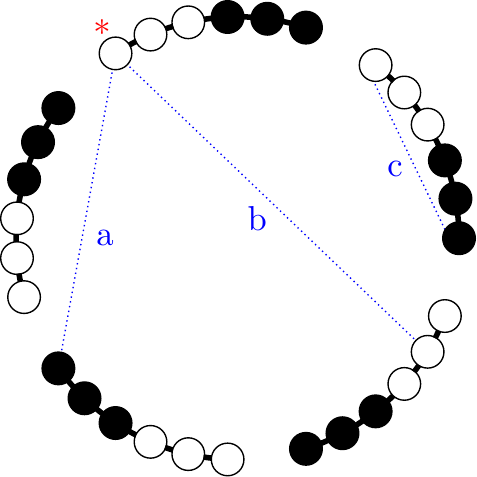}
\caption{\label{fig:cartoon} Schematic representation of the graph-based simulation model. Each circle represents a sticker on an associative polymer. Black vs. white circles correspond to two different sticker types. Covalent bonds are indicated by solid lines whereas reversible physical crosslinks are shown using blue dotted lines. The schematic shows (a) an inter-chain crosslink between white and black stickers, (b) an inter-chain crosslink between two white stickers, and (c) an intra-chain crosslink between white and black stickers. The red asterisk indicates a sticker involved in three-body interactions, which is how we model bond cooperativity.}
\end{figure}

The mean-field model assumes that each bond of type $i$-$j$ has an equivalent energy $\epsilon_{ij}$ and entropy $\log (v_{ij}/V)$. This ignores the fact that networks grow toward and past the percolation threshold by forming \emph{clusters} \cite{de1979scaling}, where a cluster is defined as a set of stickers that belong to the same polymer or distinct polymers among which physical crosslinks have formed via pairing of stickers. The overall entropy loss that accounts for \emph{intra}-cluster sticker-sticker interactions can be much smaller than the mean-field estimate \cite{Stanley1982}. Additionally, \emph{bond cooperativity}, defined as increases or decreases in the effective strengths of inter-sticker interactions that are influenced by the presence of a prior interaction, is ignored in mean-field models. We use graph-based Monte Carlo (MC) simulations of networks formed by stickers to incorporate the effects of intra-cluster interactions and model the effects due to bond cooperativity. In these simulations, stickers are nodes on a graph and the reversible physical crosslinks (bonds) are dynamic edges between nodes that engender network formation (\textbf{Figure \ref{fig:cartoon}}).

The simulation system consists of $N$ associative polymers (here $N = 100$), and each polymer contains $n_{\text{A}}$ A-type stickers and $n_{\text{B}}$ B-type stickers. The model is readily generalizable to include more than two types of stickers. Simulations are initiated from an un-networked system, and bonds between stickers are stochastically formed or broken using a Metropolis MC algorithm. In each MC step, a bond can form between a randomly chosen pair of stickers, or a randomly chosen bond can be broken. For the case of two sticker types, ignoring cooperativity, bonds of the types A-A, B-B, and A-B can form, and the interaction energy per bond is designated as $\epsilon_{ij}$ for a bond that forms between stickers $i$ and $j$. Likewise, bond breaking between stickers of type $i$ and $j$ causes a loss of energy $\epsilon_{ij}$.

Cooperativity in bond formation can be incorporated by including an energy gain (\emph{positive cooperativity}) or loss (\emph{negative cooperativity}) designated as $\epsilon_{ij|k}$. A negative value signifies positive cooperativity whereas a positive value indicates negative cooperativity associated with forming an $i$-$j$ bond given the presence of an $i$-$k$ or $j$-$k$ bond. The coordination number $\nu$ controls the maximum number of bonds each sticker can make; in the current formalism, $\nu=1$ or $\nu=2$. Cooperativity in bond formation is only realizable for $\nu>1$. In many multivalent proteins, stickers are often $\pi$-systems that include six-membered benzene rings and planar moieties such as the guanidinium group of Arg \cite{10.7554/eLife.31486}. Planar moieties can have interaction partners above or below the plane, and additional interactions are unlikely due to steric hindrance. Therefore, we set $\nu \leq 2$.

To distinguish intra-cluster vs. inter-cluster interactions, we check if a newly formed bond links two stickers that are already part of the same cluster. If they are in the same cluster, the bond is an \emph{intra}-cluster bond, and its loss in entropy ($\Delta s_{\text{intra}}$) is smaller than entropy loss for an \emph{inter}-cluster bond ($\Delta s_{\text{inter}}$); for clarity, inter-cluster bonds form between pairs of stickers from two different clusters. The losses in entropy are computed as follows:
\begin{eqnarray}
\Delta s_{\text{intra}} &=& \log \left(v_0/V_{\text{cluster}}\right), \\
\Delta s_{\text{inter}} &=& \log \left(v_0/V\right).
\end{eqnarray}
Here, $v_0$ is the bond volume, $V_{\text{cluster}}$ is the volume to which a sticker is confined within its cluster, and $V$ is the total volume of the system. For simplicity, we assume identical values of $v_0$ for all bond types and identical values of $V_{\text{cluster}}$ for all clusters.

The free energy associated with each configuration is calculated as $E - T\Delta S$, where $E$ is the summation over all energetic contributions, $\Delta S$ is the summation over all entropic contributions from the change of translational degree of freedom, and $T$ is the simulation temperature (set to 1 in this work). In each MC step, we calculate the system free energy and employ the Metropolis criterion to determine whether a proposed move (\emph{i.e.}, bond formation or destruction) will be accepted or rejected.

\begin{figure}
\includegraphics[width=86mm]{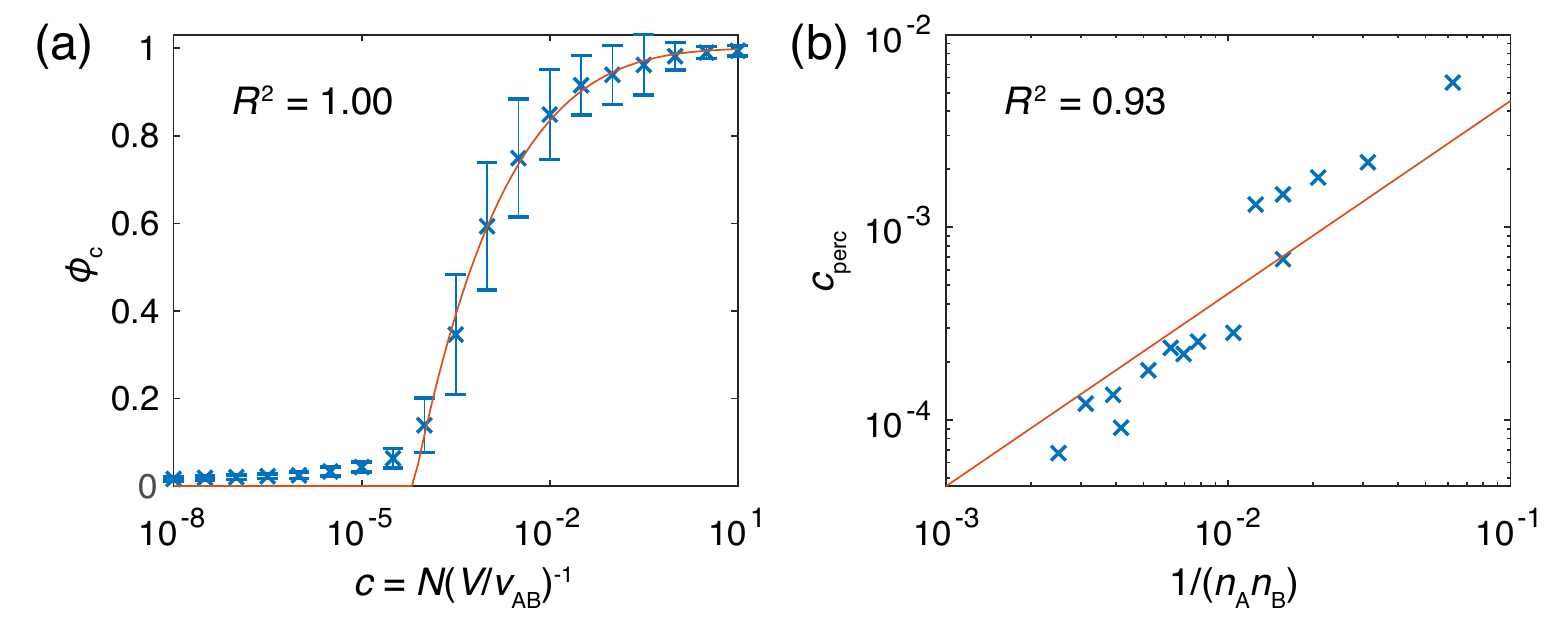}
\caption{\label{fig:SRmodel} (a) Fraction of polymers in the largest cluster ($\phi_{\text{c}}$) as a function of the concentration $c$. Here, $N = 100$, $n_{\text{A}} = 20$, $n_{\text{B}} = 20$, $\epsilon_{\text{AA}} = 10$, $\epsilon_{\text{BB}} = 10$, $\epsilon_{\text{AB}} = -3$, $\Delta s_{\text{intra}} = -20$, $k_{\text{B}}T = 1$. Error bars indicate the standard deviation across 100 independent simulations. The red line shows the curve determined by an implicit equation $1-y = e^{-xy}$, and $R^2$ is Pearson's correlation coefficient between the simulation data and the fit generated using the implicit equation. (b) Correlation between $1/(n_{\text{A}}n_{\text{B}})$ and $c_{\text{perc}}$ for the simple model where only heterotypic, inter-sticker interactions are possible. The blue crosses indicate the data points for different numbers of $n_{\text{A}}$ and $n_{\text{B}}$, and the red line shows the line of best fit $y = kx$. $R^2$ is Pearson's correlation coefficient.}
\end{figure}

Based on recent work \cite{10.7554/eLife.30294, 10.1371/journal.pcbi.1007028}, we use the fraction of polymers in the largest cluster ($\phi_{\text{c}}$) as the order parameter to quantify bond percolation, and change the reduced system volume $V/v_0$ to control the polymer concentration $c = N(V/v_0)^{-1}$ for fixed $N$. To ensure statistical robustness of our results, we performed 100 independent simulations to obtain the average value of $\phi_{\text{c}}$ for each polymer concentration. A typical profile for $\phi_{\text{c}}$ as a function of $c$ is shown in \textbf{Figure \ref{fig:SRmodel}(a)}. These data can be fit to a curve described by an implicit equation $1-y(x) = e^{-xy(x)}$, which quantifies the size of the giant component ($y$) as a function of the mean degree ($x$) in an Erd\"os-R\'enyi random graph \cite{janson2011random,latora2017complex}. The concentration threshold at which the largest component emerges ($x = 1$) is used as a proxy for the percolation threshold $c_{\text{perc}}$. Note that clusters of non-negligible sizes can form even for $x < 1$. These may be thought of as \emph{proto}-condensates that grow into networked condensates, a feature that has recently been highlighted for multivalent proteins that drive pyrenoid formation \cite{Xu:2020aa}.

We tested the accuracy of the graph-based MC simulations by assessing the ability to reproduce mean-field predictions (\textbf{Equation \ref{eq:MFT}}) for the case where only heterotypic inter-chain interactions are allowed: $\epsilon_{\text{AA}} = 10$, $\epsilon_{\text{BB}} = 10$, and $\epsilon_{\text{AB}} = -3$. Also, the intra-cluster interactions are prohibited by setting $\Delta s_{\text{intra}} = -20$. To eliminate bond cooperativity, $\nu$ is set to 1. We tested all possible combinations of $n_{\text{A}} = 4, 8, 12, 16, 20$ and $n_{\text{B}} = 4, 8, 12, 16, 20$, and the values we obtain for $c_{\text{perc}}$ from the simulations correlate well with $1/(n_{\text{A}}n_{\text{B}})$ in accordance with the mean-field model (\textbf{Figure \ref{fig:SRmodel}(b)}).

\section{Effects of intra-cluster interactions and bond cooperativity}
We deployed the graph-based MC simulations to understand how a polymer made up of two types of stickers might give rise to the observation that $c_{\text{sat}}$ of full-length FUS ends up being two orders of magnitude lower than the N-terminal PLD. For this, we compare percolation thresholds of two systems, $\text{A}_{10}$ and $\text{A}_{10} \text{B}_{10}$, which we use as phenomenological mimics of the FUS PLD and the FL FUS, respectively. Differences due to intra-cluster interactions are accounted for by setting $\Delta s_{\text{intra}} = -1$. We also query the effects of bond cooperativity by setting $\nu$ to two. We investigate the space spanned by two parameters, $\epsilon_{\text{AB}}$ and $\epsilon_{\text{AA}|\text{B}}$, while $\epsilon_{\text{AA}} = -2$, $\epsilon_{\text{BB}} = 5$, and all other terms are set to zero. As before, we use $\Delta s_{\text{inter}} = \log (v_0 / V)$ to titrate polymer concentration.

\begin{figure}
\includegraphics[width=86mm]{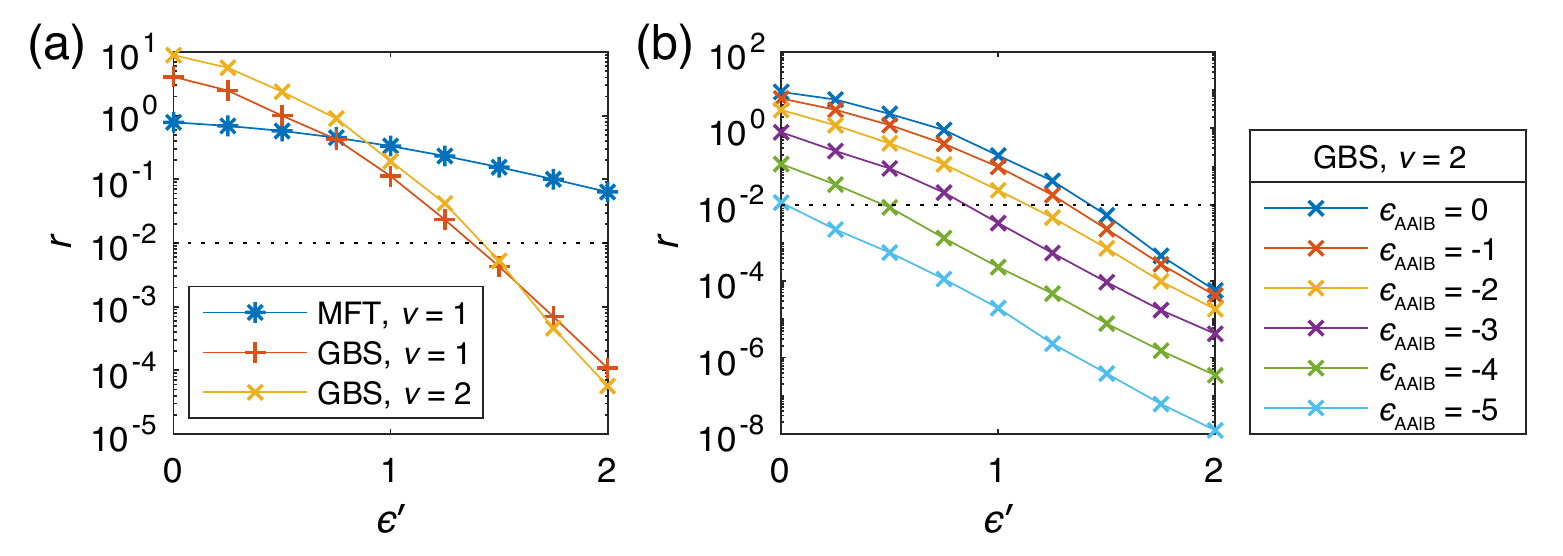}
\caption{\label{fig:params} Comparison of the variation of $r$ with $\epsilon'$. (a) Mean-field theory (MFT) and two different graph-based simulations (GBS) with $\nu = 1$ and 2. $\epsilon_{\text{AA}|\text{B}} = 0$ for all three cases. (b) Graph-based simulations with $\nu = 2$ and distinct values of $\epsilon_{\text{AA}|\text{B}}$ (legend).}
\end{figure}

We define $r$ as a measure of the relative percolation threshold, $r = c_{\text{perc}}(\text{A}_{10} \text{B}_{10}) / c_{\text{perc}}(\text{A}_{10})$, and $\epsilon'$ as the relative energetic advantage of heterotypic interactions compared to homotypic interactions, $\epsilon' = \epsilon_{\text{AB}}/\epsilon_{\text{AA}}$. \textbf{Figure \ref{fig:params}(a)} shows how $r$ varies with $\epsilon'$ when $\epsilon_{\text{AA}|\text{B}} = 0$. In the mean-field model, the $\text{A}_{10}\text{B}_{10}$ system cannot have a higher percolation threshold than $\text{A}_{10}$, irrespective of the value of $\epsilon'$. In direct contrast, inclusion of intra-cluster interactions allows for $r >1$ for small $\epsilon'$. Even though A-B interactions do not make significant energetic contributions for small $\epsilon'$, B stickers consume A stickers via intra-cluster interactions. This is an entropically driven process and reduces the effective number of available A stickers for inter-cluster interactions. Hence, for weak A-B interactions ($\epsilon' < 1$), intra-cluster interactions are favored over inter-cluster interactions, causing a clear deviation from the mean-field model. For $\epsilon' > 1$, stronger interactions between A-B stickers cause a shift between intra- and inter-cluster interactions, leading to lower values of $r$ vis-\'{a}-vis the mean-field model. \textbf{Figure \ref{fig:params}(a)} also quantifies the effect of setting $\nu=2$ for $\epsilon_{\text{AA}|\text{B}} = 0$. Although this allows more crosslinks among different chains and contributes to lowered percolation thresholds (\textbf{Figure S1}), the qualitative trends for the variation of $r$ with $\epsilon'$ are similar to the case of $\nu=1$. Overall, the graph-based simulations show that \emph{intra-cluster} interactions change the effective number of available stickers for inter-cluster interactions. The extent of this change is determined by relative sticker strengths.

Next, we tested the effects of \emph{bond cooperativity} on the variation of $r$ with $\epsilon'$. This analysis is motivated by solubility data for amino acids \cite{AUTON2007397}. These data suggest that Tyr-Tyr interactions are orders of magnitude stronger than Arg-Arg interactions. It is also known that cations can enhance $\pi$-$\pi$ interactions \cite{Mahadevi:2016aa}. Accordingly, we propose that the formation of an Arg-Tyr interaction can impact the likelihood of incorporating an additional Tyr into the network via a Tyr-Tyr crosslink to a Tyr-Arg pair. How might such effects alter bond percolation transitions of model associative polymers with A and B stickers? To answer this question, we queried the effects of changing $\epsilon_{\text{AA}|\text{B}}$. In this scenario, A-A bond formation is aided by pre-existing A-B bonds, and the strengths of these interactions are governed by the magnitude of $\epsilon_{\text{AA}|\text{B}}$. A reference value of $r=10^{-2}$ is achieved for $\epsilon' > 1$ when $\epsilon_{\text{AA}|\text{B}} = 0$; however, increasing positive bond cooperativity causes a significant lowering of $r$ (\textbf{Figure \ref{fig:params}(b)}). As an illustration, for $\epsilon_{\text{AA}|\text{B}} = -5$, the $\epsilon'$ required to achieve $r = 10^{-2}$ is almost zero, indicating that A-B bonds are no longer used for inter-cluster interactions, but instead they serve the sole purpose of strengthening A-A bonds. Positive bond cooperativity shifts the balance toward inter-cluster interactions, leading to a lowering of the percolation threshold for the $\text{A}_{10}\text{B}_{10}$ system when compared to the $\text{A}_{10}$ system.

In the context of our model, there are many combinations of $\epsilon_{\text{AB}}$ and $\epsilon'$ that lower the percolation threshold of $\text{A}_{10}\text{B}_{10}$ versus $\text{A}_{10}$. At one extreme, B stickers directly interact with A stickers ($\epsilon_{\text{AB}} \approx -3$ and $\epsilon_{\text{AA}|\text{B}} = 0$); and at the other extreme, their only role is to be enhancers of A-A interactions ($\epsilon_{\text{AB}} = 0$ and $\epsilon_{\text{AA}|\text{B}} \approx -5$). Accordingly, we quantify the impact of changing the sticker valence $n_{\text{A}}$ and $n_{\text{B}}$ for different interaction paradigms. We investigated four sets of parameters: $\{\epsilon_{\text{AB}}, \epsilon_{\text{AA}|\text{B}}\} = \{-3, 0\}, \{-2, -2.5\}, \{-1, -4\}, \{0, -5\}$ because these parameter sets lead to $r \approx 10^{-2}$ (\textbf{Figure \ref{fig:params}(b)}).

\begin{figure}
\includegraphics[width=86mm]{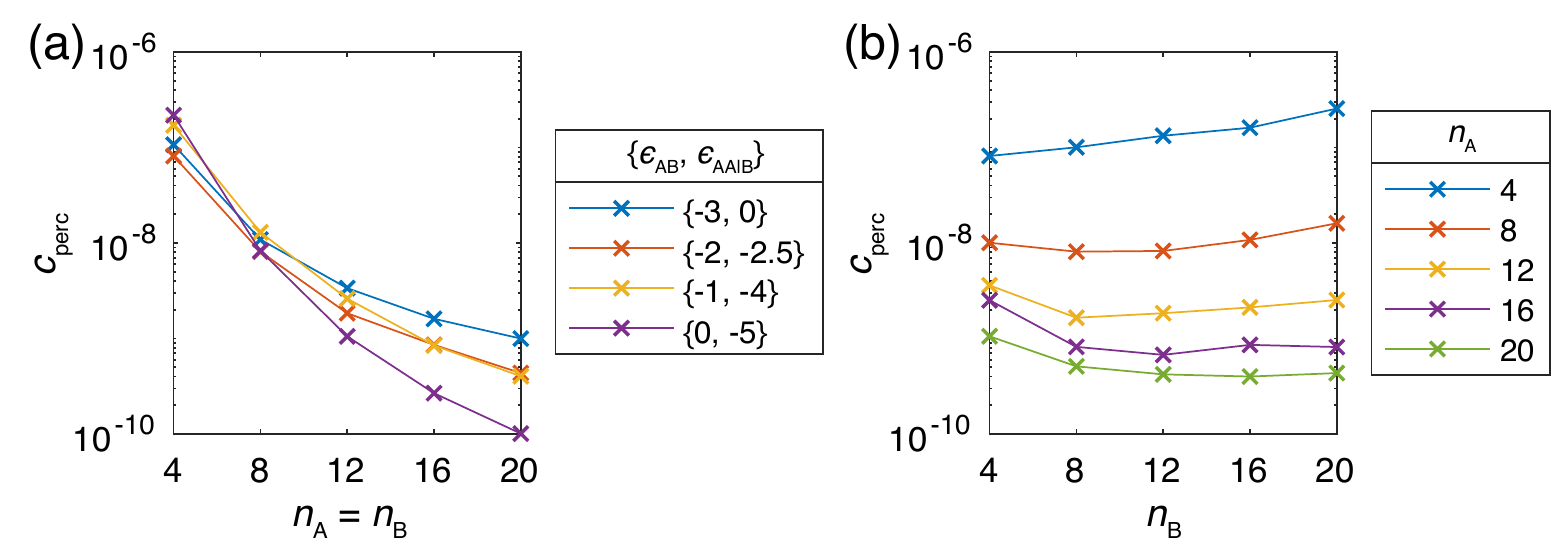}
\caption{\label{fig:stickerno} Sticker numbers and percolation concentrations. (a) Variation of $c_{\text{perc}}$ as a function of valence for the symmetric system where $n_{\text{A}} = n_{\text{B}}$. Each curve corresponds to a different combination of $\epsilon_{\text{AB}}$ and $\epsilon_{\text{AA}|\text{B}}$ (see legend). (b) Variation of $c_{\text{perc}}$ as a function of $n_{\text{B}}$ for systems with different $n_{\text{A}}$ (see legend). For all systems, $\epsilon_{\text{AB}} = -1$ and $\epsilon_{\text{AA}|\text{B}} = -4$.}
\end{figure}

The mean-field model (\textbf{Equation \ref{eq:MFT}}) predicts that $c_{\text{perc}}$ decreases when either $n_{\text{A}}$ or $n_{\text{B}}$ increases. This behavior is preserved in graph-based simulations when $n_{\text{A}}$ and $n_{\text{B}}$ simultaneously increase (\textbf{Figure \ref{fig:stickerno}(a)}). However, for fixed $n_{\text{A}}$, the absence or weakening of heterotypic A-B interactions leads to an increase in the percolation threshold by an order of magnitude as $n_{\text{B}}$ increases, and this cannot be compensated by increasing bond cooperativity. Further, the inclusion of bond cooperativity leads to changes in how $c_{\text{perc}}$ varies with $n_{\text{B}}$ for fixed $n_{\text{A}}$. These are analyzed for $\epsilon' = 0.5$ ($\epsilon_{\text{AB}} = -1$) and $\epsilon_{\text{AA}|\text{B}} = -4$ (\textbf{Figure S2} shows other cases). As shown in \textbf{Figure \ref{fig:stickerno}(b)}, $c_{\text{perc}}$ increases with increasing $n_{\text{B}}$ for $n_{\text{A}} = 4$; this derives from the paucity of A stickers. A surprising result is that $c_{\text{perc}}$ can also vary non-monotonically with increasing $n_{\text{B}}$ as is illustrated for $n_{\text{A}} = 8$ and $n_{\text{A}} = 12$. Specifically, as $n_{\text{B}}$ increases beyond $n_{\text{A}}$, the decrease of $c_{\text{perc}}$ is replaced by an increase of $c_{\text{perc}}$ as $n_{\text{B}}$ increases. These results arise because A-A interactions as opposed to A-B interactions are required for strong bond cooperativity. However, when $n_{\text{B}} > n_{\text{A}}$, more of the A stickers are consumed in entropically favored intra-cluster interactions with B stickers. This hinders inter-cluster interactions thereby inhibiting growth of the network and causing an increase in $c_{\text{perc}}$ vis-\'{a}-vis the situation where bond cooperativity is ignored. The implication is that positive bond cooperativity will enhance bond percolation when $n_{\text{B}} < n_{\text{A}}$ and weaken bond percolation when $n_{\text{B}} > n_{\text{A}}$. 

\section{Summary}
We have generalized extant mean-field models for bond percolation to quantify the effects of arbitrary numbers of sticker types and capture the interplay between homotypic and heterotypic sticker-sticker interactions. Using this model, we show that the joint valence of Tyr and Arg stickers combined with a small energetic advantage in Tyr-Arg interactions over Tyr-Tyr interactions is sufficient to explain experimental observations for proteins such as FUS. 

Mean-field models overestimate the entropic cost associated with forming crosslinks when clusters are already present. Improved estimates for the entropic cost, generated using graph-based MC simulations, shows that the percolation threshold for the $\text{A}_{10}$ system can actually be lower than of the $\text{A}_{10}\text{B}_{10}$ system, especially when the strengths of A-B interactions are lower than those of A-A interactions. 

We define positive bond cooperativity as the increased likelihood of forming a new $\text{A}_i$-$\text{A}_j$ bond given the prior formation of either an $\text{A}_i$-$\text{B}_k$ or $\text{A}_j$-$\text{B}_k$ bond. Strong positive bond cooperativity can fundamentally alter the dependence of $n_{\text{perc}}$ on sticker valence. In the regime where $n_{\text{B}} < n_{\text{A}}$, we predict that $c_{\text{perc}}$ decreases with increasing $n_{\text{B}}$. This behavior derives from the synergistic effects of A-B bond formation and bond cooperativity that enhances A-A interactions. However, in the regime where $n_{\text{B}} > n_{\text{A}}$, $c_{\text{perc}}$ increases with increasing $n_{\text{B}}$. In this scenario, strong positive bond cooperativity becomes refractory to bond percolation because these interactions trap the B stickers in small clusters that inhibit growth of the network. 

The effects of bond cooperativity can be unmasked by engineering mismatches between the valence of aromatic stickers in PLDs and cationic stickers in RBDs in systems such as FUS. It is also noteworthy that previous analysis of disordered proteins across the human proteome shows that $n_{\text{Arg}} \gtrsim n_{\text{Tyr}}$ \cite{WANG2018688} in most sequences. Investigating how $c_{\text{perc}}$ or $c_{\text{sat}}$ changes across disordered proteins that have similar values of $n_{\text{Tyr}}$ and differences in $n_{\text{Arg}}$ should help unmask the contributions from bond cooperativity.

\begin{acknowledgments}
This work was supported by the Basic Science Research Program through the National Research Foundation of Korea funded by the Ministry of Education (grant 2019R1A6A1A10073887 to JMC), the National Institutes of Health (grant 5R01NS056114 to RVP) and the St. Jude Research Collaborative on membraneless organelles (RVP). We thank Hyunkyu Choi, Stephan Grill, Alex Holehouse, and Christoph Zechner for helpful discussions.
\end{acknowledgments}

\bibliography{references}

\end{document}